\documentclass[epj]{svjour}

\usepackage[pdftex]{graphicx}
\usepackage[english]{babel}
\usepackage{latexsym,amsmath,verbatim}
\usepackage{color}
\usepackage{microtype}
\usepackage{hyperref}

\newcommand{\av}[1]{\langle {#1} \rangle}
\newcommand{\g}{\gamma}
\newcommand{\eps}{\varepsilon}

\newcommand{\FigPath}{./}

\begin{document}

\title{Slow relaxation dynamics and aging in random walks on activity
  driven temporal networks}

\author{Ang\'{e}lica S. Mata\inst{1,2} \and Romualdo Pastor-Satorras\inst{2}}

\institute{Departamento de F\'{\i}sica, Universidade Federal de
  Vi\c{c}osa, 36571-000, Vi\c{c}osa - MG, Brazil \and Departament de
  F\'{\i}sica i Enginyeria Nuclear, Universitat Polit\`ecnica de
  Catalunya, Campus Nord B4, 08034 Barcelona, Spain}
  
\titlerunning{Slow relaxation and aging in random walks on activity
  driven networks}

\authorrunning{A. S. Mata and R. Pastor-Satorras}

\date{Received: date / Revised version: date}

\abstract{%
  We investigate the dynamic relaxation of random walks on temporal
  networks by focusing in the recently proposed activity driven model
  [Perra \textit{et al.} Sci. Rep. srep00469 (2012)]. For realistic
  activity distributions with a power-law form, we observe the presence
  of a very slow relaxation dynamics compatible with aging effects. A
  theoretical description of this processes in achieved by means of a
  mapping to Bouchaud's trap model. The mapping highlights the profound
  difference in the dynamics of the random walks according to the value
  of the exponent $\g$ in the activity distribution.
} 

\maketitle
\section{Introduction}

The heterogeneous topology of a complex network \cite{Newman2010} can
have a very relevant impact on the properties of dynamical systems
running on top of it
\cite{dorogovtsev07:_critic_phenom,barratbook}. Already classical
studies in network science have thus shown that a heterogeneous
connectivity pattern can lead to a null percolation threshold
\cite{Cohen00,PhysRevLett.85.5468}, set a strong resilience against
random failures \cite{albert2000error}, as well as to induce a vanishing
epidemic threshold for disease propagation
\cite{Pastor-Satorras:2014aa}, indicative of a strong weakness against
the infective agents. Similar and additional remarkable effects have
been observed in a wide variety of dynamical processes, both in and out
of equilibrium (see Refs.~\cite{dorogovtsev07:_critic_phenom,barratbook}
for extensive reviews on this subject).

Such dynamical effects, originally reported for \textit{static} networks
\cite{Newman2010}, in which nodes and edges are fixed and do not change
over time, can take a different, more complex turn when one considers
the intrinsic time-varying, \textit{temporal} nature of many real
networks \cite{Holme:2011fk}. Indeed, networked systems are often not
static, but show connections which appear and disappear with some
characteristic time scales that can be of the same order of magnitud of
those ruling a dynamical process on top of the network.  Social networks
\cite{Jackson2010} represent the prototypical example of this behavior,
being defined in terms of a sequence of social contacts that are
continuously established and broken.  This mixing of time scales can
induce new phenomenology on dynamics of temporal networks, in stark
contrast with what is observed in static networks. Moreover, the
\textit{bursty} nature
\cite{Oliveira:2005fk,Onnela:2007,10.1371/journal.pone.0011596} of the
time evolution of temporal network contacts, characterized by long
stretches of inactivity, interspersed by bursts of intense activity, can
complicate the picture, inducing for example a noticeable dynamical
slowing down in dynamical processes as varied as epidemic spreading,
diffusion or synchronization
\cite{PhysRevLett.98.158702,PhysRevX.4.011041,dynnetkaski2011,Stehle:2011nx,albert2011sync}.


One of the simplest dynamical processes, although still underlying many
practical realistic applications, is the random walk
\cite{WeissRandomWalk}. Even in this simplest of cases, a time-varying
substrate can induce very noticeable differences with respect to the
behavior expected in static networks.
\cite{PhysRevE.85.056115,perra_random_2012,ribeiro_quantifying_2013,hoffmann_generalized_2012}. Particularly
relevant in this sense is the analysis of the random walk behavior in
\emph{activity driven} networks \cite{2012arXiv1203.5351P}, a class of
social temporal network models based on the observation that the
establishment of social contacts is driven by the activity of
individuals, prompting them to interacte with their peers with different
levels of intensity, and on the empirical measurement of heterogeneous
levels of activity $a$ across different datasets, activity which is
found to be distributed according to a power law form, $F(a) \sim
a^{-\g}$ \cite{2012arXiv1203.5351P}. The analytic study of the random
walk in this class of models, focusing on steady state, large time
properties, points out the striking differences imposed by a
time-varying topology, with respect to a static one
\cite{perra_random_2012}. In the present paper we extend the study of
these differences, by considering the time evolution of the dynamics
towards the steady state. Using a combination of analytic arguments and
numerical simulations, we show that random walks on activity driven
networks exhibit a slow relaxation to their steady state. The time scale
of the relaxation is inversely proportional to a parameter $\eps$,
measuring the smallest activity in the network. In the limit $\eps\to0$,
we found evidence of \textit{aging} behavior in the random walk
relaxation \cite{Henkel2}, characterized by a breaking of time
translation invariances for time scales smaller than $\eps^{-1}$. By
means of a mapping to Bouchaud's trap model
\cite{bouchaud2012,monthus1996} we show that, for $\g <1$, random walks
on activity driven networks exhibit simple aging, characterized by a
unique relevant time scale. On the other hand, the case $\g>1$
corresponds to a more complex picture, with several competing
characteristic time scales.

We have organized our manuscript as follows: In
Sec.~\ref{sec:activ-driv-netw} we recall the definition of the activity
driven model. Sec.~\ref{sec:activ-rand-walks} defines the continuous
time implementation of directed random walks on activity driven networks
and derives the analytic steady state solution for the occupation
probability in terms of a generalized Montroll-Weiss equation. In
Sec.~\ref{sec:ageing-random-walks} we present numerical evidence for the
aging behavior of the occupation probability $P(a, t)$, defined as the
probability that the walker is in a node of activity $a$ at time $t$. In
Sec.~\ref{sec:mapp-bouch-trap} we develop the mapping of the random walk
process to Bouchaud's trap model. The mapping suggests additional
quantities characterizing the aging behavior of the system, which are
numerically analyzed.  Our conclusions are finally presented in
Sec.~\ref{sec:conclusions}.

\section{The activity driven network model}
\label{sec:activ-driv-netw}

The activity driven network model
\cite{2012arXiv1203.5351P,starnini_topological_2013} is defined as
follows: $N$ nodes (individuals) in the network are endowed with an
activity $a_i \in [\eps, 1]$, extracted randomly from an activity
distribution $F(a)$. Every time step $\Delta t = 1/N$, an agent $i$ is
chosen uniformly at random. With probability $a_i$, the agent becomes
active and generates $m$ links that are connected to $m$ other agents,
chosen uniformly at random. Those links last for a period of time
$\Delta t$ (i.e. are erased at the next time step). Time is updated by $t
\to t + \Delta t$. For simplicity, we will consider in the following
$m=1$. The topological properties of the integrated network at time $t$
(i.e. the network in which nodes $i$ and $j$ are connected if there has
ever been a connection between them at any time $t' \leq t$) have been
studied in Ref.~\cite{starnini_topological_2013}, obtaining as a main
result that the integrated degree distribution at time $t$, $P_t(k)$,
scales in the large $t$ limit as the activity distribution, i.e.
\begin{equation}
  P_t(k) \sim t^{-1} F\left( \frac{k}{t} - \av{a} \right).
  \label{eq:10}
\end{equation}
Empirical measurements report activity distributions in real temporal
networks exhibiting long tails of the form $F(a) \sim a^{-\gamma}$
\cite{2012arXiv1203.5351P}.  This expression thus relates in a simple
way the functional form of the activity distribution and the degree
distribution of the integrated network at time $t$, and allows to
explain the scale-free form of the latter observed in social networks
\cite{newmancitations01,newman_scientific_2001}.

In this paper we will consider activity distributions with this
power-law form. The range of values of the $a$ is restricted to $a \in
[\eps, 1]$, where a minimum activity $\eps$ is set to avoid divergencies
close to zero. The normalized form of the distribution thus depends on
the value of $\gamma$:
\begin{equation}
  \label{eq:4}
  F(a) = \frac{1-\g}{1-\eps^{1-\g}} \; a^{-\g}.
\end{equation} 
As we will see later, the parameter $\eps$ will play a significant role
in the analysis of random walks on activity driven networks.

\section{Random walks on activity driven networks}
\label{sec:activ-rand-walks}

The dynamics of a random walk on activity driven networks is defined as
follows
\cite{perra_random_2012,ribeiro_quantifying_2013,hoffmann_generalized_2012}:
A walker arriving at a node $j$ at time $t$ remains on it until an edge
is created joining $i$ and other node $j$ at a subsequent time $t' >
t$. The walker then jumps instantaneously to node $j$ and waits there
until an edge departing from it is created. To simplify calculations,
here we will focus on \textit{activated random walks}: a walker can
leave node $i$ only when $i$ becomes active and creates an edge pointing
at another node \cite{perra_random_2012}. Once the walker has arrived to
node $i$, it must wait there until $i$ creates a new connection. Since
$i$ creates new edges with constant probability $a_i$ per unit time, the
walker will remain trapped in $i$ for a number of time steps $n$ given
by the exponential distribution, $\psi_i(n) = \frac{a_i}{N}
\left(1-\frac{a_1}{N} \right)^{n-1}$, independently of the time of the
last activation of $i$. In the limit of large $N$ we can take the
continuous time limit and define a \textit{waiting time} $\tau = n/N$,
which is given by a local waiting time distribution
\begin{equation}
  \psi_{a_i}(\tau) = a_i e^{-a_i \tau}.
  \label{eq:5}
\end{equation}
That is, the dynamics of hopping from one node to another, follows in
time a Poisson process with a rate $a_i$ that depends on node $i$.

The dynamics of activated random walks under this restriction is
particularly easy to implement in continuous time: Considering that the
walker is at vertex $i$ with activity $a_i$ at time $t$, it hops to a
randomly selected node $j$ and time is updated as $t \to t + \tau$,
where $\tau$ is a random variable extracted from the distribution
Eq.~(\ref{eq:5}).  This continuous time implementation has the
additional benefit of not restricting the maximum possible value of the
activity $a$, which can be now considered as a probability rate. With
this definition, a directed random walk on an activity driven netwok can
be directly mapped to a continuous time random walk (CTRW) on a fully
connected network in which each node has a different distribution of
waiting times $\psi_i(\tau)$ \cite{WeissRandomWalk}.

\subsection{Steady state solution}
\label{sec:steady-state-solut}

The time evolution of the activated random walk on activity driven
networks can be studied by means of the generalized Montroll-Weiss
equation approach \cite{WeissRandomWalk,hoffmann_generalized_2012}. For
a Poissonian waiting time distribution, as in Eq.~(\ref{eq:5}), the
occupation probability $P(i,t)$ of finding the walker in node $i$ at
time $t$ fulfills the exact equation \cite{hoffmann_generalized_2012}
\begin{equation}
  \frac{d P(i,t)}{d t} = - \Lambda_i P(i,t) +  \sum_j \lambda_{ij}
  P(i,t) ,
  \label{eq:gmw}
\end{equation}
where $ \lambda_{ij}$ is the probability per unit time that the walker
jumps from node $j$ to node $i$, and $\Lambda_j = \sum_i \lambda_{ij}$
is the probability per unit time that the walker at $j$ leaves this
node. For the activated random walk on activity driven networks, we
obviously have $\lambda_{ij} = a_j / N$ and $\Lambda_j =
a_j$. Eq.~(\ref{eq:gmw}) then reads
\begin{equation}
  \frac{d P(i,t)}{d t} = - a_i P(i,t) +  \frac{1}{N} \sum_j a_j P(j,t).
  \label{eq:6}
\end{equation}
We can obtained an effective equation for the probability $P(a,t)$ that
the walker is in a node of activity $a$ at time $t$ by performing a
coarse-graining of Eq.~(\ref{eq:6}), in which we define $P(a,t) =
\sum_{i \in \mathcal{V}(a)} P(i, t)$, where $\mathcal{V}(a)$ is the set
of nodes with activity $a$, with an average size $N_a = N
F(a)$. Applying this definition on Eq.~(\ref{eq:6}), and rearranging the
summation over $j$, we obtain
\begin{equation}
   \frac{d P(a, t)}{d t} = -a P(a, t) + F(a) \sum_{a'} a' P(a', t).
   \label{eq:7}
\end{equation}
From Eq.~(\ref{eq:7}) it is straightforward to obtain the steady state
solution $\lim_{t \to \infty} P(a, t) \equiv P_\infty(a)$ by imposing
$\frac{d P(a,t)}{d t} = 0$, obtaining
\begin{equation}
  P_\infty(a) =  \frac{F(a)}{a} \sum_{a'} a' P_\infty(a') \equiv
  \frac{1}{\av{a^{-1}}} \frac{F(a)}{a},
  \label{eq:8}
\end{equation}
where in the last term we have applied the normalization condition
$\sum_a P_\infty(a) = 1$, thus recovering the result obtained in
Ref.~\cite{perra_random_2012}.

\section{Slow relaxation dynamics}
\label{sec:ageing-random-walks}

\begin{figure}[t]
   \begin{center}
     \includegraphics*[width=8cm]{\FigPath/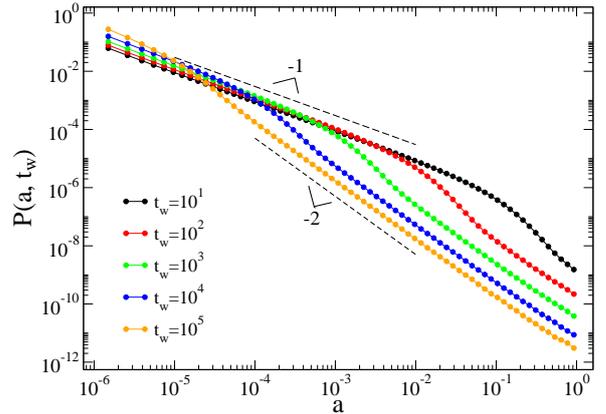}
     \caption{Evolution towards equilibrium of the occupation
       probability $P(a, t_w)$, measured after a time $t_w$, in activity
       driven networks with activity distribution $F(a) \sim a^{-\g}$,
       with $\g =1$ and minimum activity $\eps = 10^{-6}$.  For small
       $t_w$, the occupation probability is proportional to the activity
       distribution, $P(a,t_w \to 0) \sim F(a) \sim a^{-1}$; for large
       $t_w$, it saturates to the steady state form $P(a,t_w \to \infty)
       = P_\infty(a) \sim F(a)/a \sim a^{-2}$. Network size $N = 5
       \times 10^6$.}
     \label{fig:ageing_bare}
   \end{center}
\end{figure}

Eq.~(\ref{eq:7}) yields information about the occupation probability of
nodes with activity $a$ at large times, Eq.~(\ref{eq:8}), expression
whose accuracy has been checked
numerically~~\cite{perra_random_2012}. From it, however, it is hard to
extract information about the time evolution of the process, and in
particular, about the time scales of the relaxation to the steady
state. We explore this issue by means of numerical simulations. Thus, in
Fig.~\ref{fig:ageing_bare} we plot the occupation probability $P(a,
t_w)$ of nodes of activity $a$, measured after letting the walker evolve
for a time $t_w$. 
As Fig.~\ref{fig:ageing_bare} shows, the occupation probability exhibits
a very slow relaxation from a state $P(a, t_w\to0) \sim F(a)$ at short
times to the equilibrium state, $P_\infty(a) \sim F(a)/a$, at large
times, see Eq.~(\ref{eq:8}). As a function of $a$ for fixed $t_w$, this
relaxation translates in a crossover between both scaling regimes at a
crossover activity $a_c(t_w)$ which is a decreasing function of $t_w$.

We can understand the origin of this crossover by the following
argument~\cite{PhysRevE.80.020102}: The average time $\tau_a$ to exit
from a node with activity $a$ is
\begin{equation}
  \tau_a = \int_0^\infty \tau \psi_a(\tau) = \frac{1}{a}.
  \label{eq:2}
\end{equation}
A walker initially at a node of activity $a$, is expected to remain
there for any time smaller than $\tau_a$. Since the smallest activity in
the network is $\eps$, for $t_w > \eps^{-1}$ the walker has had the
chance to explore (and scape from) all nodes in the network, and
therefore we expect to find it in the steady state.  For any arbitrary
time $t_w < \eps^{-1}$, one can thus consider that all nodes with
activity $a$ such that $\tau_a<t_w$ (large $a$) will have had time to
relax and reach the steady state, while nodes with activity fulfilling
$\tau_a>t_w$ (small $a$) will not have relaxed.  We thus see that the
crossover activity fulfills $\tau_{a_c} \sim t_w$ or, from
Eq.~(\ref{eq:2}), $a_c(t_w) \sim t_w^{-1}$. The previous argument
suggests therefore the following scaling form for the whole occupation
probability:
\begin{equation}
  P(a, t_w) = t_w \; \mathcal{P}(a \; t_w),
  \label{eq:3}
\end{equation}
where $\mathcal{P}(z)$ is a scaling function satisfying 
\begin{equation}
  \mathcal{P}(z) \sim \left\{
    \begin{array}{cl}
       z^{-\g} & \mathrm{for}\; z \ll 1\\
       z^{-\g-1} & \mathrm{for}\; z \gg 1
    \end{array}
    \right.
    \label{eq:19}
\end{equation} 
This scaling regime is expected to hold for times $t_w < \eps^{-1}$,
i.e. before the full relaxation to the steady state.

\begin{figure}[t]
   \begin{center}
     \includegraphics*[width=7cm]{\FigPath/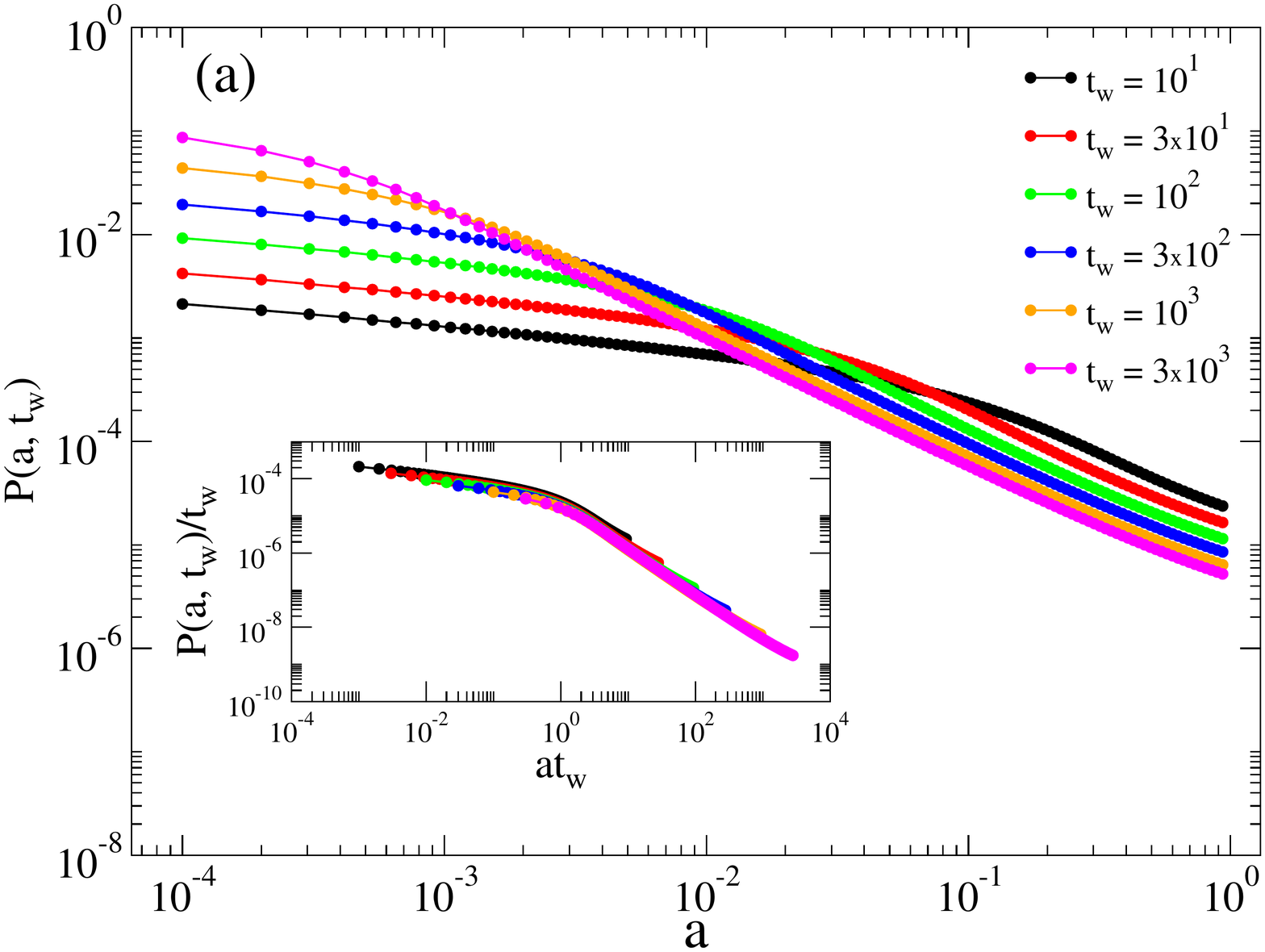}
     \includegraphics*[width=7cm]{\FigPath/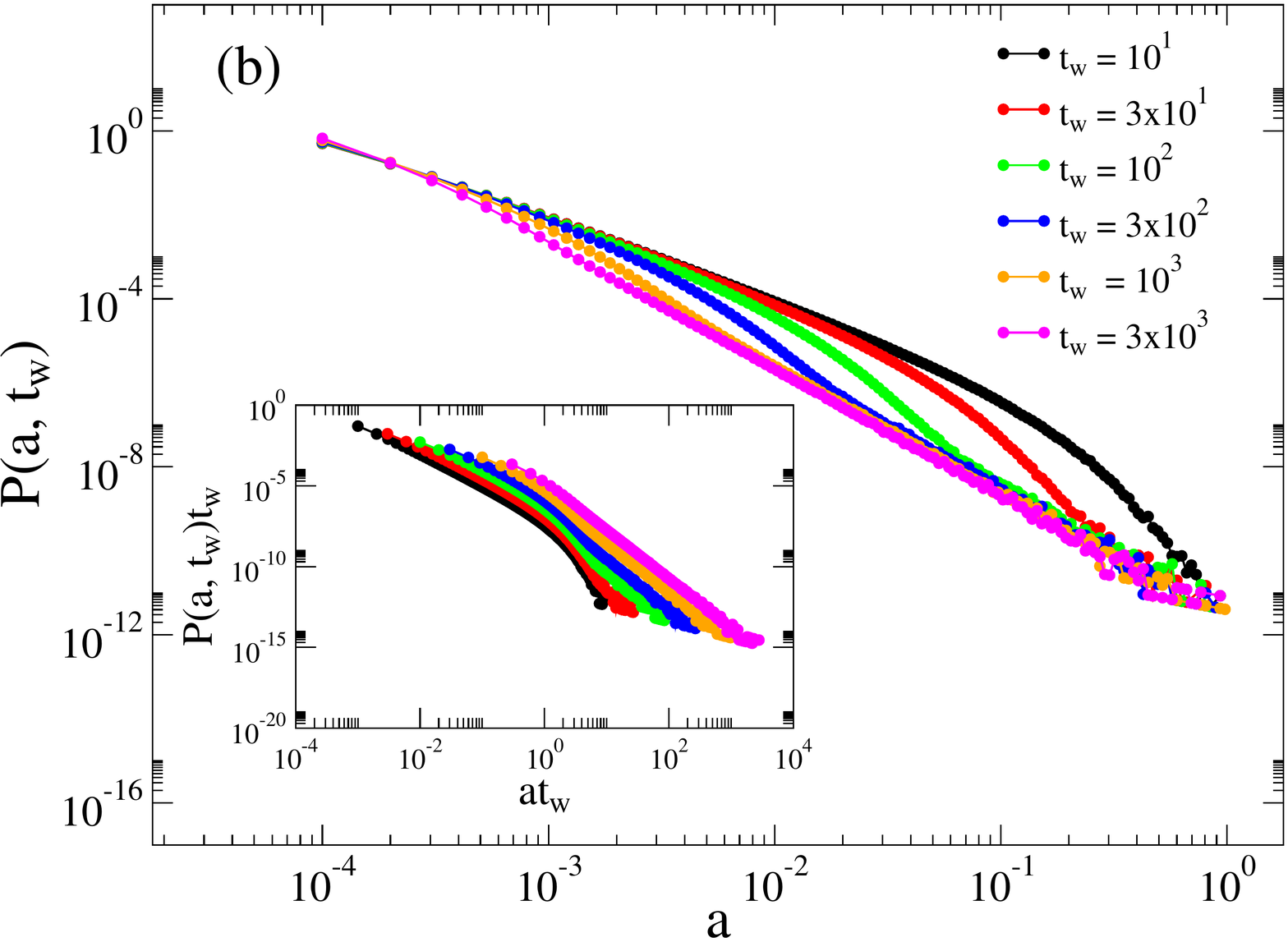}
     \caption{Occupation probability $P(a,t_w)$ as a function of the
       activity $a$ at different times $t_w$. Data refer to
       activity-driven networks with $N = 5\times 10^6$, $\eps =
       10^{−4}$, and (a) $\gamma=0.25$, (b) $\gamma=2.00$. Insets: Data
       collapse according to the scaling form Eq.~(\ref{eq:3}).}
     \label{fig:ageing_scaling}
   \end{center}
\end{figure}

In Fig.~\ref{fig:ageing_scaling} we check the scaling form in
Eq.~(\ref{eq:3}) for activated random walks in activity driven networks
with power-law activity distribution.
For values of $\gamma<1$, Fig.~\ref{fig:ageing_scaling}(a), we observe
that the scaling form of the occupation probability is perfectly
fulfilled for all times $t_w < \eps^{-1}$. Surprisingly, however, the
scaling form fares quite badly for $\gamma>1$, performing increasingly
worst for larger values of $\gamma$.

\section{Mapping to Bouchaud's trap model and aging behavior}
\label{sec:mapp-bouch-trap}

The radical difference in behavior of the random walk for $\g$ larger or
smaller than $1$ can be understood in terms of a mapping to the
well-known Bouchaud's trap model \cite{bouchaud2012,monthus1996} for
glassy behavior (see also
\cite{PhysRevE.80.020102,1742-5468-2011-03-P03032}). The trap model is
defined in terms of a phase space consisting on $N$ \textit{traps}, each
one with a depth $E_i$, $i=1,\ldots,N$, extracted randomly from the
probability distribution $\rho(E)$. The dynamics of the model proceeds
by a succession of jumps between the traps, ruled by the temperature of
the system, $T$. At this temperature, the system remains in a trap of
depth $E$ a random time $\tau$ distributed according to a Poisson
process with rate $\tau_E^{-1} = \exp(-E/T) / \tau_0$, where $\tau_0$ is
a microscopic time scale that can be arbitrarily set equal to $1$. After
this time, the system jumps to a randomly chosen trap. Since all traps
are equivalent, the probability that the system lands on a trap of depth
$E$ after a jump is $\rho(E)$. The average time spent in any trap is
thus
\begin{equation}
  \av{\tau} = \int \rho(E) \tau_E \; dE.
  \label{eq:26}
\end{equation}
The trap model exhibits a phase transition between a high temperature
phase, where $\tau$ is finite, to a low temperature, glassy state
characterized by very slow relaxation dynamics and aging behavior
\cite{Henkel2}, when $\av{\tau}$ diverges. This transition takes place
at a finite temperature $T_0$ for a depth probability distribution of
the form $\rho(E) = \exp(-E/T_0)$ \cite{bouchaud2012,monthus1996}, in
which case the distribution of trapping times takes the form
\begin{equation}
  P(\tau) \sim \tau^{-1-T/T_0}
  \label{eq:21}
\end{equation}
For $T<T_0$, the exponent in the power-law in Eq.~(\ref{eq:21}) is
smaller than $2$ and thus leads to an infinite average trapping time.

Activated random walkers can be easily mapped to the trap model by
noticing the mapping between the respective jumping rates, that is
\begin{equation}
  a = \tau_E^{-1} = \exp(-E/T),
  \label{eq:11}
\end{equation}
form where we obtain the equivalence between depth and activity
\begin{equation}
  E = - T \ln (a),
  \label{eq:12}
\end{equation}
with a range of variation $E \in [0, T \ln(\eps^{-1})]$.
The relation between the activity distribution and the depth
distribution is given by $\rho(E) \; dE = F(a) \; da$, leading to
\begin{equation}
  \rho(E) = \frac{e^{-E/T}}{T} F\left(e^{-E/T}\right). 
\label{eq:13}
\end{equation}
Assuming now an activity with a power-law distribution as is
Eq.~\eqref{eq:4}, we obtain
\begin{equation}
  \rho(E)   \sim \frac{1}{T} \exp \left[ -(1-\g)E/T \right].
  \label{eq:14}
\end{equation}
Let us consider separately the different possible values of $\g$.

\subsection{Case $\gamma < 1$}

From Eq.~\eqref{eq:14}, the mapping to the trap model makes perfect
sense for $\g <1$. In this case, $\rho(E)$ is a decreasing function of
the depth $E$, corresponding to the presence of many shallow traps and a
few deep ones. Deep traps represent rare long trapping events that
eventually dominate the dynamics below the glass transition and induce a
very slow relaxation and aging behavior
\cite{bouchaud2012,monthus1996}. In this case, however, temperature
plays no role in the equivalent dynamics of the activated random
walkers, as it can be absorbed in a change of variables $\bar{E} = E /
T$, with a range of variation $\bar{E} \in [0, \ln(\eps^{-1})]$.  From
the point of view of the mapping to the trap model, the activated random
walker is a system in a fully glass state, corresponding to a infinite
glass transition temperature. This can easily be seen by looking at the
average waiting time distribution, i.e
\begin{eqnarray}
  \label{eq:20}
  \psi(\tau) &=& \sum_a F(a) a e^{-a \tau} \simeq   \frac{1-\g}{1-\eps^{1-\g}}
  \int_\eps^\infty    a^{1-\g} e^{-a \tau} da = \nonumber \\
  &\simeq &   \tau^{-2+\g} e^{- \eps \tau},
  \label{eq:27}
\end{eqnarray} 
where we have made an expansion for $\tau < \eps^{-1}$.  The exponent in
Eq.~(\ref{eq:27}) is always smaller than $2$, indicative of an infinite
glass transition temperature. The average trapping time is thus
modulated by the exponential factor,
diverging when $\eps \to 0$, i.e. when the upper cut-off of the
associated energy tends to infinity.

\begin{figure}[t]
   \begin{center}
     \includegraphics*[width=8cm]{\FigPath/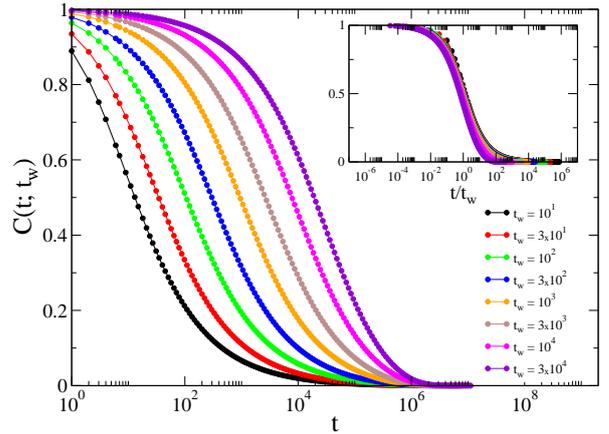}
     \caption{Two-time correlation function for random walks on activity
       driven networks with a power-law activity distribution. Data
       refers to $N=5\times10^6$, $\g = 0.50$ and $\eps
       =10^{-6}$. Inset: Scaling plot corresponding to the form in
       Eq.~(\ref{eq:16}).}
     \label{fig:correls_gamma_small}
   \end{center}
\end{figure}

This analogy allows to explore in the random walk problem other features
of the glassy dynamics of the trap model, in particular aging effects
\cite{Henkel2}. Aging effects are here usually measured by looking at
the two-time correlation function $C(t; t_w)$, between the states of the
system at times $t_w$ and $t + t_w$. This correlation function, which is
defined as the average probability that the system in a given trap at
time $t_w$ has not performed a jump at time $t+t_w$, fulfills in the
trap model the scaling relation
\begin{equation}
  C(t; t_w) = \mathcal{C}\left( \frac{t}{t_w} \right),
  \label{eq:16}
\end{equation}
corresponding to the so-called "simple" aging
\cite{bouchaud2012,monthus1996}. 
This scaling can be simply deduced for the random walk on activity
driven networks: Since the jumping dynamics is Poissonian in every node,
the probability of not leaving a node with activity $a$ in a time
interval $t$ is $e^{-a t}$. We thus can write
\begin{eqnarray}
  C(t; t_w) &=& \int_\eps^1 P(a, t_w) e^{-a t} \; da 
  =\int_\eps^1 t_w \mathcal{P}(a \;t_w) e^{-a t} \; da \nonumber \\
  &=& \int_{t_w \eps}^{t_w} \mathcal{P}(z) e^{-z t/t_w} \; dz ,
\end{eqnarray}
where we have used the scaling relation Eq.~(\ref{eq:3}) for $P(a, t_w)$
and performed a change of variable. For large $z$, the upper limit of
the integral can be safely set equal to infinity, due to the exponential
cut-off. In the limit of small $z$, we have $\mathcal{P}(z) \sim
z^{-\g}$, and its integral also converges for $t_w \eps$ small. Thus, in
the double limit $1 \ll t_w \ll \eps^{-1}$, we have
\begin{equation}
   C(t ; t_w) \simeq \int_{0}^{\infty} \mathcal{P}(z) e^{-z t/t_w} \; dz,
\end{equation}
recovering the scaling relation for the correlation Eq.~(\ref{eq:16}),
which is expected to hold for waiting times $t_w \ll \eps^{-1}$. In
Fig.~\ref{fig:correls_gamma_small} we show that, for $\gamma<1$, the
scaling of the correlations is very well fulfilled in the random walk
process.

\begin{figure}[t]
   \begin{center}
     \includegraphics*[width=8cm]{\FigPath/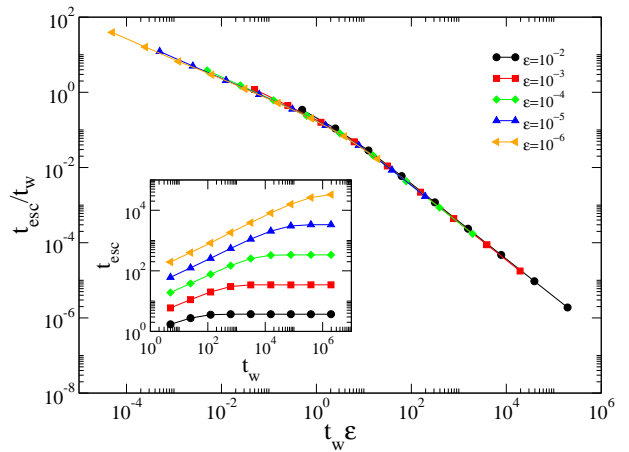}
     \caption{Scaling plot of the average scape time according to
       Eq.~(\ref{eq:18}) for $\gamma=0.50$.  Inset: raw data.  Data
       corresponds to a network size $N = 10^5$.}
     \label{fig:escape_time_gamma_small}
   \end{center}
\end{figure}

Additional information on the aging properties of the system can be
gathered by looking at the average escape time $t_\mathrm{esc}(t_w)$
that a walker at a given node at time $t_w$ requires to escape from it
\cite{PhysRevE.80.020102}. In
Fig.~\ref{fig:escape_time_gamma_small}(inset) we plot the curves of the
escape time as a function of $t_w$, for different values of $\eps$. As
we can see, the escape time is a increasing function of $t_w$ for small
values of $t_w$, indicating another of the typical features of aging
systems, namely a breaking of scale invariance translation.  Due to its
Poissonian nature, the average time to leave a node with activity $a$ is
$\tau_a = 1/a$.  Therefore, we can write $t_\mathrm{esc}(t_w) = \int da
\;P(a, t_w) / a$. Applying the scaling relation Eq.~(\ref{eq:3}) and
performing a change of variables, we have
\begin{equation}
   t_\mathrm{esc}(t_w) = t_w \int_{\eps t_w}^{t_w}
   \frac{\mathcal{P}(z)}{z} \; dz.
\end{equation}
For large $t_w$, the upper limit of the integral is not singular, and we
can set it to infinity. We are therefore led to the scaling form
\begin{equation}
   t_\mathrm{esc}(t_w) \simeq t_w \mathcal{F}(t_w \; \eps),
   \label{eq:18}
\end{equation}
which is valid for $t_w \ll \eps^{-1}$.

We check this theoretical predictions by means of a data collapse
analysis, plotting $ t_\mathrm{esc}(t_w)/t_w$ as a function of $t_w
\eps$. In Fig.~\ref{fig:escape_time_gamma_small}(main) we plot the
result, showing an excellent agreement with the prediction.

\subsection{Case $\gamma > 1$}

\begin{figure}[t]
   \begin{center}
     \includegraphics*[width=8cm]{\FigPath/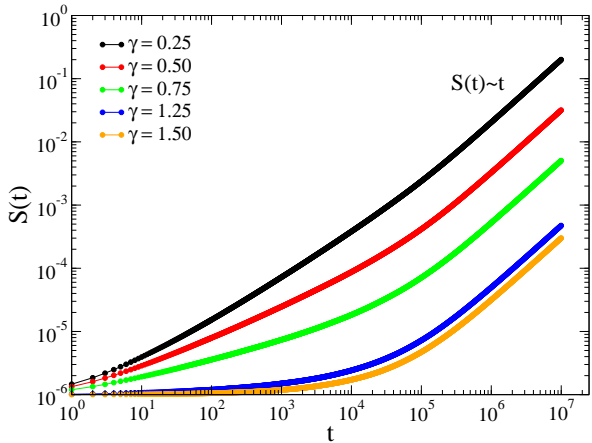}
     \caption{Coverage as a function of time. For $\g<1$, dashed lines
       represent the predicted behavior $\sim t^{1-\g}$ from
       Eq.~\eqref{eq:25}. We use $N = 10^6$ and $\eps = 10^{-5}$.} 
     \label{fig:coverage}
   \end{center}
\end{figure}

For the case $\g >1$, Eq.~\eqref{eq:14} indicates that the mapping to
the trap model is not physical. For this range of $\g$ values, the
density of traps \textit{increases} with depth, meaning that very deep
traps are much more probable than shallow ones. Long time trapping is
thus not a rare event, but the norm of the system. This implies a
qualitatively different, much slower dynamics than for the case $\g <
1$. This effect can be simply seen by looking at the coverage $S(t)$ of
the random walk as a function of time, defined as the average fraction
of different nodes that a walker has visited up to time $t$, see
Fig.~\ref{fig:coverage}.
For very large $t > \eps^{-1}$, the system has reached the steady state,
and the walker jumps from any node at an average increment time $\Delta
t = \av{\tau}$, where the average waiting time $\av{\tau}$ is defined as
\begin{equation}
  \label{eq:22}
  \av{\tau} = \int_{\eps}^1 F(a) \tau_a \; da = \av{a^{-1}} = 
  \frac{1-\g}{\g} \frac{\eps^{-\g} -1}{1-\eps^{1-\g}},
\end{equation}
where we have used Eqs.~\eqref{eq:22} and~\eqref{eq:4}. In this limit of
large $t$, and since walker jumps to randomly chosen new nodes, we
expect 
\begin{equation}
  \label{eq:23}
  S(t) \sim  \frac{t}{\av{a^{-1}}}, \quad t \gg 
  \eps^{-1}. 
\end{equation}
For $\g<1$ and short times, we observe a power law increase of the
coverage, $S(t) \sim t^{\alpha}$. In the case $\g>1$, on the other hand
the initial growth is extremely slow, slower numerically than
logarithmical, in the region $t < \eps^{-1}$, where the dynamics is
dominated by an exceedingly large number of deep traps. After this very
slow regime, the linear, steady state behavior is recovered.  We can
explain the exponent of the initial growth of $S(t)$ for $\g<1$ by the
following argument: At any time $t \ll \eps^{-1}$, the random walker
will have explored in average nodes with activity restricted to $a >
t^{-1}$, since it will not have had time to escape deeper traps. In this
case, the average inverse activity of those nodes explored will depend
on time as
\begin{equation}
  \label{eq:24}
  \av{a^{-1}}_t \sim \int_{t^{-1}}^1 \frac{F(a)}{a} \; da \sim t^{\g}.
\end{equation}
We therefore estimate a coverage
\begin{equation}
  \label{eq:25}
  S(t) \sim \frac{t}{\av{a^{-1}}_t} \sim t^{1-\g},
\end{equation}
with an exponent $\alpha = 1-\g$ which is a decreasing function of $\g$,
in agreement with the results in Fig.~\ref{fig:coverage}.  

The anomalous density of deep traps for $\g>1$ has a strong effect in
the scaling of the occupation probability. In
Fig.~\ref{fig:ageing_scaling}(b) we have plotted the function $P(a,
t_w)$ for $\g = 2.0$. From the rescaled plot, it is evident that a
simple scaling behavior such as the one represented in Eq.~(\ref{eq:3})
is doomed to fail. In particular, as we can observe, the occupation
probability for small values of $a$ is apparently independent of
$t_w$. This fact is in stark contrast with the behavior for $\gamma<1$,
Fig.~\ref{fig:ageing_scaling}(a), in which the value of $P(a, t_w)$ for
small $a$ increases for increasing $t_w$, as expected from
Eq.~(\ref{eq:3}). This phenomenology is again due to the anomalous
abundance of deep traps in the region $\g>1$: thus, while at $t_w$ the
walker has had the opportunity to explore and escape from nodes with
activity $a > t_w^{-1}$, the fraction of such nodes is very small for
large values of $\g$. Indeed, denoting by $\phi(t_w)$ the fraction of
such nodes, we have
\begin{equation}
  \label{eq:28}
  \phi(t_w) = \int_{t_w^{-1}}^1 F(a) = \frac{1-t_w^{\g-1}}{1-\eps^{1-\g}}.
\end{equation}
Denoting by $\phi^>(t_w)$ (resp.  $\phi^<(t_w)$) the fraction
corresponding to $\g>1$ (resp. $\g<1$), we have, in the limit of small
$\eps$,
\begin{equation}
  \label{eq:29}
  \phi^>(t_w) \sim (t_w \eps)^{\g-1}, \quad \phi^<(t_w) \sim 1- t_w^{\g-1}.
\end{equation}
I.e. the fraction of visited nodes up to time $t_w$ grows much faster
for $\g<1$ than in the opposite case.

\begin{figure}[t]
   \begin{center}
     \includegraphics*[width=8cm]{\FigPath/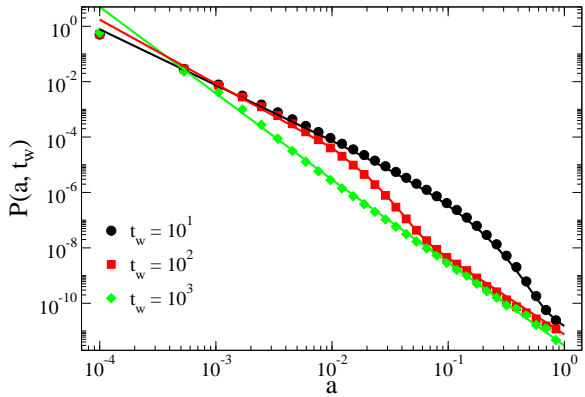}
     \caption{Evolution of the occupation probability $P(a,t)$. Data
       refer to activity-driven network with $N = 5\times 10^6$, $\eps =
       10^{−4}$, $\gamma=2.00$, and different waiting times $t_w$.
       Solid lines are non-linear regressions using Eq.~(\ref{eq:31}).}
     \label{fig:occupation_theoretical_gamma_large}
   \end{center}
\end{figure}

While the scaling relation Eq.~(\ref{eq:3}) is invalid for $\g >1$, we
can still understand the functional form of the occupation probability
by means of the following argument: Let us consider a walker at time
$t_w$. Initially, the walker starts at a randomly chosen node with
activity $a_0$. With probability $e^{a_0 t_w}$, the walker does not move
from its initial position, and therefore it contributes to the initial
activity $a_0$. With the complementary probability $1- e^{a_0 t_w}$, the
walker has performed one or more jumps. Assuming that after its first
jump it already reaches the steady state $F(a) / [a \av{a^{-a}}]$, we
have the average occupation probability
\begin{equation}
  \label{eq:30}
  P(a, t_w) = F(a) e^{-a t_w} + \frac{F(a)}{a \av{a^{-1}}} \left [1 -
      \int_\eps^1 e^{-a_0 t_w} F(a_0) \right]. \nonumber
\end{equation}
For the activity distribution given by Eq.~(\ref{eq:4}), we have
  \begin{eqnarray}
    \label{eq:31}
    P(a, t_w) &=& F(a) e^{-a t_w} + \frac{F(a)}{a \av{a^{-1}}} \times
    \nonumber \\
    &\times& \left [ 1 -
      \frac{1-\g}{1-\eps^{1-\g}}t^{-1+\g} \Gamma(1-\g, t_w \eps) \right],
  \end{eqnarray}
  where in the integral we have extended the upper limit up to infinity,
  and $\Gamma(a,z)$ is the incomplete Gamma function
  \cite{abramovitz}. In
  Fig.~\ref{fig:occupation_theoretical_gamma_large} we plot the
  numerical data for $\g=2$, compared with the theoretical prediction in
  Eq.~(\ref{eq:31}).
As we can see, except for very small values of the activity $a$, the fit
of the theoretical prediction with the numerical results is
excellent. The mixing of time scales in Eq.~(\ref{eq:31}), namely
$a^{-1}$ and $\eps^{-1}$, allows to understand the failure of the simple
scaling ansatz performed to derive Eq.~(\ref{eq:3}). 

\begin{figure}[t]
   \begin{center}
     \includegraphics*[width=8cm]{\FigPath/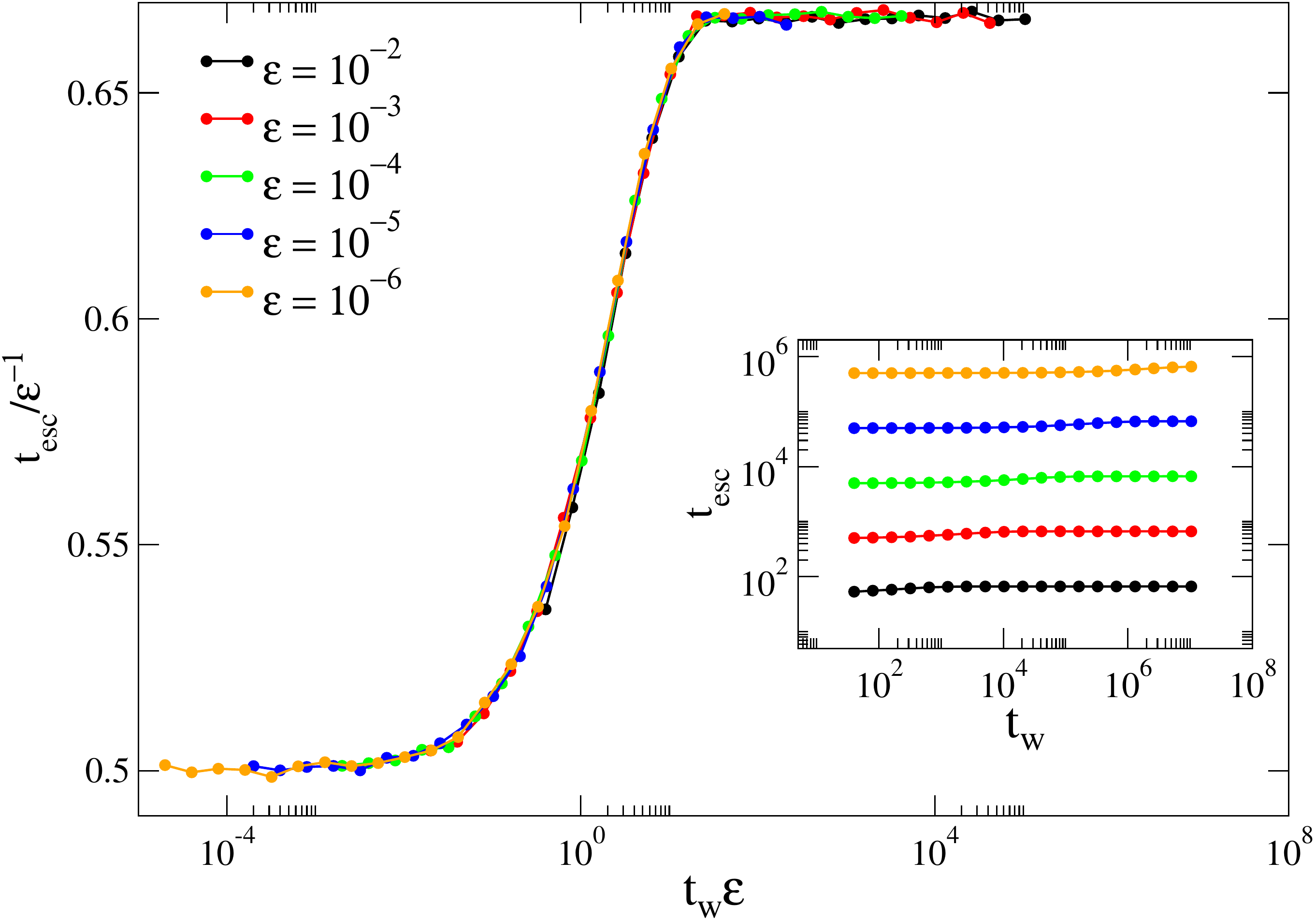}
     \caption{Scaling plot of the average scape time according to
       Eq.~(\ref{eq:32}) for $\gamma=2.00$.  Inset: raw data.  Data
       corresponds to a network size $N = 10^5$.}
     \label{fig:escape_time_gamma_large}
   \end{center}
\end{figure}

Finally, let us consider the average escape time $t_{esc}(t_w)$ for $\g
> 1$. The lack of simple scaling evidenced in Eq.~(\ref{eq:31})
indicates that the simple collapse predicted in Eq.~(\ref{eq:18}) cannot
be correct, see Fig.~\ref{fig:escape_time_gamma_large}(inset).
We can recover however some form of scaling law for this
quantity. Considering a fixed value of $\eps$, the plateau at large
$t_w$ is given by the escape time at the stationary state,
$t_{esc}^{ss}$, which is independent of $t_w$ and, from
Eq.~(\ref{eq:8}), given by $t_{esc}^{ss} = \int P_\infty(a)/a \;da =
\av{a^{-2}}/\av{a^{-1}} \sim \eps^{-1}$, far all $\g$. On the other
hand, for small $t_w$, $t_{esc}$ is dominated by the deepest traps, and
it starts only to increase when the walker has had time to explore a
finite fraction of the network, that is, when time is larger than the
average trapping time $\av{\tau}$, which, from Eq.~\eqref{eq:22} is
proportional to $\eps^{-1}$ for $\g>1$. This reasoning suggest a scaling
behavior for the escape time of the form
\begin{equation}
  \label{eq:32}
  t_{esc}(t_w) = \eps^{-1} \mathcal{G}(t_w \eps).
\end{equation}
This scaling form in checked in
Fig.~\ref{fig:escape_time_gamma_large}(main), where we can see that it
is quite well satisfied. Incidentally, this scaling form can also be
cast in the form valid for $\g<1$, Eq.~(\ref{eq:18}), by simply defining
$\mathcal{F}(z) \equiv \mathcal{G}(z)/z$.

\section{Conclusions}
\label{sec:conclusions}

In this paper we have investigated the temporal relaxation of the
simplest dynamical process, namely the random walk, on the class of
activity driven temporal networks.  We have focused in particular in the
case of activated random walks, in which a walker can only leave a node
when the latter becomes active. By means of a combination of analytic
calculations and numerical experiments, we have shown that, for networks
with a power law distribution of activity, the random walk experiences a
very slow relaxation towards its steady state. The speed of this
relaxation is mainly controlled by the parameter $\eps$, bounding the
smallest activity of any node. In the limit of small $\eps\to0$, the
dynamics exhibits aging behavior, characterized additionally by a
breaking of time translation symmetry. The aging properties of the
random walk are studied by examining different quantities usually
applied to characterize aging in glassy systems. Crucially, the aging
properties of the random walk depend on the exponent $\g$ in the
activity distribution. For $\g<1$, the random walk exhibits a relaxation
compatible with "simple" aging, with a unique characteristic time scale
given by the average escape time from the least active node,
$\eps^{-1}$. In this regime, simple scaling forms for the two-time
correlation function and the average escape time can be worked out,
starting from a scaling ansatz for the occupation probability $P(a,
t_w)$. For $\g>1$, on the other hand, the picture is more complex, with
a scaling ruled by several characteristic time scales. This different
behavior according to $\g$ can be understood by means of a mapping to
Bouchaud's trap model: The case $\g>1$ corresponds in this case to an
unphysical representation of the trap model, in which there is a
majority of deep traps that induce an extraordinary slow relaxation
dynamics.

The results obtained here indicate that, apart form the slowing down
already reported in dynamical systems on temporal networks, more complex
effects, such as aging, can also be observed. In the present case, in
which the temporal network substrate is the activity driven model, aging
emerges as the result of the mixing of Poisson activation processes with
widely different time scales.  A more complex phenomenology is to be
expected in real temporal networks with non Poissonian, bursty
activation rates.

\section*{Acknowledgments}
We thank A. Barrat for very helpful comments and discussions. This work
has been supported by the CAPES under project No.5511-13-5.  RP-S
acknowledges financial support from the Spanish MINECO, under projects
No. FIS2010-21781-C02-01 and FIS2013-47282-C2-2, and EC FET-Proactive
Project MULTIPLEX (Grant No. 317532).


\end{document}